\documentclass[aps,prl,twocolumn,preprintnumbers,floatfix,showpacs,superscriptaddress]{revtex4-1}
\usepackage[dvips]{graphics}
\usepackage{graphicx}
\usepackage{dcolumn}
\usepackage{bm}
\usepackage{epstopdf}
\usepackage{amsmath}
\usepackage{amsfonts}
\usepackage{amssymb}
\usepackage{xcolor}
\usepackage{subfigure}
\usepackage{braket}
\usepackage{commath}
\usepackage[colorlinks=true,linkcolor=blue,anchorcolor=red,citecolor=blue,urlcolor=blue]{hyperref}

\providecommand{\U}[1]{\protect\rule{.1in}{.1in}}

\begin{document}

\title{Thermoelectric Generation of Orbital Magnetization in Metals}

\author{Cong Xiao}
\thanks{These authors contributed equally to this work.}
\affiliation{Department of Physics, The University of Texas at Austin, Austin, Texas 78712, USA}

\author{Huiying Liu}
\thanks{These authors contributed equally to this work.}
\affiliation{Research Laboratory for Quantum Materials, Singapore University of Technology and Design, Singapore 487372, Singapore}

\author{Jianzhou Zhao}
\affiliation{Co-Innovation Center for New Energetic Materials, Southwest University of Science and Technology, Mianyang 621010,
China}
\affiliation{Research Laboratory for Quantum Materials, Singapore University of Technology and Design, Singapore 487372, Singapore}

\author{Shengyuan A. Yang}
\affiliation{Research Laboratory for Quantum Materials, Singapore University of Technology and Design, Singapore 487372, Singapore}

\author{Qian Niu}
\affiliation{Department of Physics, The University of Texas at Austin, Austin, Texas 78712, USA}

\begin{abstract}
We propose an orbital magnetothermal effect wherein a temperature gradient
generates an orbital magnetization (OM) for Bloch electrons, and we present a unified theory for electrically and thermally induced OM, valid for both metals and insulators. We reveal that there exists an intrinsic response of OM, for which the susceptibilities are completely determined by the band geometric quantities such as interband Berry connections, interband orbital moments, and the quantum metric. The theory can be readily
combined with first-principles calculations to study real
materials. As an example, we calculate the OM response in CrI$_{3}$
bilayers, where the intrinsic contribution dominates. The temperature
scaling of intrinsic and extrinsic responses, the effect of phonon drag, and the phonon angular momentum contribution to OM are discussed.
\end{abstract}

\maketitle

Orbital magnetization (OM) is an important fundamental property of solids,
yet its theoretical calculation is notoriously difficult, made possible only
relatively recently. The main problem is that the magnetic dipole operator $%
\bm r\times \bm v$ is ill defined for Bloch basis, which are the eigenstates
for extended periodic systems. Several approaches, such as semiclassical~\cite{Xiao2005},
Wannier function~\cite{Resta2005}, and thermodynamic approaches~\cite{Shi2007}, have been developed to
circumvent this problem, and succeeded in establishing a formula of OM for a
system at \emph{equilibrium}.

OM may also be produced by external driving forces, e.g., through the
magnetoelectric effect~\cite{Vanderbilt2010,Murakami2015,Zhong2016}. Indeed,
recent experiments reported signals of pronounced OM generated by applied
electric field in doped monolayer MoS$_{2}$ and twisted bilayer
graphene~\cite{Mak2017,Gordon2019,Lee2019}, which are essentially
two-dimensional (2D) metals.

This field-generated OM has distinct symmetry requirement from the
equilibrium OM. As shown in Table~\ref{tab:Basic}, while the equilibrium OM
requires the unperturbed system to have broken time-reversal ($\mathcal{T}$)
symmetry, the linear-order electrically generated OM, i.e.,
\begin{equation}
\Delta M_{i}=\chi _{ij}E_{j},
\end{equation}%
requires the inversion symmetry ($\mathcal{P}$) to be broken. Here $\chi
_{ij}$ is defined to be the susceptibility tensor, and summation over
repeated Cartesian indices is implied henceforth. Interestingly, one
observes that if the system simultaneously breaks $\mathcal{T}$, there could
exist an \textquotedblleft intrinsic" contribution, meaning that the
corresponding susceptibility $\chi ^{\text{int}}$ is determined solely by
the band structure of the material, independent of the scattering
(which gives the extrinsic contribution with $\chi ^{\text{ext}}$).
Furthermore, if the combined $\mathcal{PT}$ symmetry is respected, $\Delta %
\bm M$ will only have the intrinsic contribution (see Table~\ref{tab:Basic}).

Despite the exciting experimental discovery and the symmetry argument of its
existence, so far, we do not have a coherent theory for the electrically
generated OM in metals. We stress the metallic state here, because for
insulators, one may derive a theory by extending the previous approaches,
but for metals, which are pertinent to many experiments, these approaches do
not work. For instance, there is problem in defining localized Wannier
functions in metals \cite{Vanderbilt2010,Resta2005,Moore2010,Lee2011}, and in the presence of
current flow (indicating an out-of-equilibrium system) the thermodynamic
approach~\cite{Xiao2005,Shi2007} is not applicable. This poses an outstanding challenge in condensed matter
physics.

\begin{table}[tbp]
\caption{Symmetry requirements of the equilibrium OM and the induced OM $%
\Delta M$ (generated by the electric field or by temperature gradient). The
last two columns indicate the temperature dependence of each contribution in
$\Delta M$ in the low-$T$ and the high-$T$ regimes.}
\label{tab:Basic}%
\begin{tabular}{cccccc}
\hline\hline
OM & $\ \ \ \ \mathcal{P}$ & \ \ $\ \ \mathcal{T}$ & \ \ \ $\ \mathcal{PT}$
& \ \ low-$T$ & \ \ \ \ high-$T$ \\ \hline
&  &  &  &  &  \\
equilibrium OM & \ \ \ \checkmark & \ \ $\times$ & \ \ \ $\times$ &  &  \\
&  &  &  &  &  \\
intrinsic $\Delta M$ & \ \ \ $\times$ & \ \ $\times$ & \ \ \ \checkmark & $\
\ \ \ \ \ \sim T$ & $\ \ \ \ \ \ \sim T$ \\
&  &  &  &  &  \\
extrinsic $\Delta M$ & \ \ \ $\times$ & \ \ \checkmark & \ \ \ \ $\times$ & $%
\ \ \ \ \ \ \sim T$ & $\ \ \ \ \ \ \sim T^{0}$ \\ \hline\hline
\end{tabular}%
\end{table}

Meanwhile, the temperature gradient $\nabla T$ shares the same symmetry as
the $\bm E$ field, hence, from the symmetry perspective, there should also
exist OM generated by $\nabla T$,
\begin{equation}
\Delta M_{i}=\alpha _{ij}(-\partial _{j}T),
\end{equation}%
with the same characters as in Table~\ref{tab:Basic}. Such an effect, which
may be termed as the orbital magnetothermal effect, has not been explored
before. For this effect, besides the problems in treating the OM in metals,
there is additional complication in dealing with $\nabla T$: as a
statistical force, it does not directly enter into the single-particle
Hamiltonian as a perturbation.

In this work, we predict the orbital magnetothermal effect, and we present a
unified theory for the thermally and electrically generated OM in 2D systems, applicable for both metals and insulators. We show
that the induced OM can be extracted from the magnetization current, which
in turn can be derived from a semiclassical theory integrating the recently
developed field variational \cite{Dong2020} and second-order wave packet~%
\cite{Gao2014} methods. Particularly, we obtain elegant formulas for the
intrinsic response coefficients $\chi^{\text{int}}$ and $\alpha ^{\text{int}%
} $, expressed by band geometric quantities such as interband Berry
connections, interband orbital moments, and the quantum metric.
Intriguingly, we find that $\chi^{\text{int}}$ and $\alpha^{\text{int}}$
fulfill the generalized Mott relation, thus measuring one allows us to also
extract the other. Our theory can be readily combined with first-principles
density-functional-theory (DFT) calculations to study real materials. As an
example, we calculate the OM response in CrI$_{3}$ bilayers, where the
intrinsic contribution dominates due to the $\mathcal{PT}$ symmetry.

{\color{blue}\textit{Approach}.} The starting point of our approach is the
intrinsic connection between OM $\bm M$ and the \emph{local} current density
$\bm j$. It is well known in electromagnetism that $\bm j$ of a nonuniform
steady state consists of two parts: the transport current and the
magnetization current:
\begin{equation}
\boldsymbol{j}=\boldsymbol{j}^{\text{tr}}+\boldsymbol{j}^{\text{mag}},\text{
\ \ }\boldsymbol{j}^{\text{mag}}=\nabla\times\bm M\mathbf{.}  \label{current}
\end{equation}
The magnetization current $\bm j^{\text{mag}}$ does not contribute to the
net current flow through a sample \cite{Cooper1997}, thus what one measures
in transport experiment is $\bm j^{\text{tr}}$. Here, the key observation is
that for 2D systems, OM is a pseudoscalar, as $\bm M$ only has the
out-of-plane ($z$) component. Consequently, the OM can be completely
determined from $\bm j^{\text{mag}}$ through the second equation of (\ref%
{current})~\cite{note-uncertainty}. Thus, the task can be reduced to
identifying $\bm j^{\text{mag}}$ in the local current $\bm j$.

There are still three obstacles for this task. First, we need an unambiguous
separation of $\bm j^{\text{mag}}$ from $\bm j^{\text{tr}}$ in $\bm j$. This
is nontrivial, because terms in $\bm j^{\text{tr}}$ may also involve spatial
derivatives (like $\partial _{i}T$). Fortunately, this difficulty is solved
in our recent work, by the trick of a fictitious inhomogeneous field
implemented in the semiclassical theory. The inhomogeneous field, assumed to
be a vector, $\bm w(\bm r)$, gives a spatial dependence of the electron
wave packet state and energy, which helps to distinguish the $\bm j^{\text{%
mag}}$ contribution. The specific form of $\bm w$ does not matter, and it is
set to zero at the end of the calculation. In \cite{Xiao2020EM}, this method
has successfully reproduced the formulas for the equilibrium OM.

Second, since we are looking for the induced OM $\Delta \bm M$ which is of
linear order in the perturbation, $\bm j^{\text{mag}}$ must be evaluated to
the second order (one order is from the spatial derivative of $\bm w$). This
means that, to calculate the current we must employ a semiclassical theory
with second order accuracy. Fortunately, such a framework is developed
in our recent work~\cite{Gao2014} and has found successful applications in various
nonlinear effects~\cite{Gao2015,Gao2017}.

Third, after clarifying the above two points, calculating the
electric-field induced OM becomes straightforward. However, we still need a way to incorporate the statistical forces (such as $\nabla T$
and $\nabla \mu $, with $\mu $ the chemical potential). This is achieved by generalizing
the recently developed field variational approach~\cite{Dong2020} to
incorporate the second-order semiclassical dynamics, which gives a unified treatment of both electric field and statistical forces.

In the Supplemental Material~\cite{supp}, we present detailed derivations
based on the wave packet action $S$, where the local current can be formally
expressed as (set $\hbar =1$)
\begin{equation}
\boldsymbol{j}\left( \boldsymbol{r}\right) =\int \left[ d\boldsymbol{k}_{c}%
\right] d\boldsymbol{r}_{c}\mathcal{D}f_{\text{tot}}\frac{\delta S}{\delta
\boldsymbol{A}\left( \boldsymbol{r}\right) }\Big|_{\boldsymbol{A}\left(
\boldsymbol{r}\right) \rightarrow 0}.  \label{key-link}
\end{equation}%
Here, $(\bm r_{c},\bm k_{c})$ are the center of the electron wave packet in
phase space, $\left[ d\boldsymbol{k}_{c}\right] \equiv \sum_{n}d\boldsymbol{k%
}_{c}/\left( 2\pi \right) ^{2}$ with $n$ the band index, $\mathcal{D}$ is
the modified phase space measure~\cite{Xiao2010}, $f_{\text{tot}}$ is the
occupation function, and $\bm A$ is the vector potential,
an auxiliary field which is set to zero at the end of the derivation.

Taking into account the fictitious inhomogeneous field, the variation yields
(the subscripts $c$ are dropped hereafter and we omit the band index here)
\begin{equation}
\boldsymbol{j}=e\int \left[ d\boldsymbol{k}\right] \mathcal{D}f_{\text{tot}}%
\boldsymbol{\dot{r}}+\boldsymbol{\nabla }\times \int \left[ d\boldsymbol{k}%
\right] \mathcal{D}f_{\text{tot}}(\boldsymbol{\tilde{m}}+\dot{k}_{j}\partial
_{\bm B}a_{j}^{B}),  \label{j-loc}
\end{equation}%
where $(\dot{\bm r},\dot{\bm k})$ are given by the second-order equations of
motion%
\begin{align}
\dot{\boldsymbol{r}}=& \ \partial _{\boldsymbol{k}}\tilde{\varepsilon}-\dot{%
\boldsymbol{k}}\times \mathbf{\tilde{\Omega}}-\mathbf{\tilde{\Omega}}_{%
\boldsymbol{kr}}\cdot \dot{\boldsymbol{r}},  \notag \\
\dot{\boldsymbol{k}}=& \ e\boldsymbol{E}-\partial _{\boldsymbol{r}}\tilde{%
\varepsilon}+\mathbf{\tilde{\Omega}}_{\boldsymbol{rk}}\cdot \dot{\boldsymbol{%
k}}.  \label{EOM}
\end{align}%
Here, $e(<0)$ is the electron charge, $\boldsymbol{\tilde{m}}$ is the
orbital magnetic moment, $\tilde{\varepsilon}$ is the wave packet energy, $%
\mathbf{\tilde{\Omega}}$ is the momentum space Berry curvature, $\mathbf{%
\tilde{\Omega}}_{\boldsymbol{kr}}=-\mathbf{\tilde{\Omega}}_{\boldsymbol{rk}}$
is the phase space Berry curvature, and $\mathcal{D}=1+\text{Tr}\,\mathbf{\tilde{%
\Omega}}_{\boldsymbol{kr}}$. The tilde in these symbols indicates that they
include corrections from the external fields. The detailed expressions for
these quantities do not concern us here, and can be found in the
Supplemental Material \cite{supp}. $\bm a^{B}$ is known as the field-induced
positional shift \cite{Gao2014}, representing the linear correction to the $%
\boldsymbol{k}$-space Berry connection by the magnetic field $\bm B$, hence
the term $\partial _{\bm B}a_{j}^{B}$ in (\ref{j-loc}) is independent of $%
\bm B$. The importance of this term to the induced OM will be shown shortly.

We make the following observations on the result in Eq.~(\ref{j-loc}).
First, to distinguish intrinsic and extrinsic contributions, we write $f_{%
\text{tot}}=f_{0}+\delta f$ in (\ref{j-loc}), with $f_{0}$ the equilibrium
Fermi distribution and $\delta f$ the off-equilibrium part. Then, the
intrinsic contribution of the induced OM $\Delta \bm M^{\text{int}}$ will
contain terms with only $f_{0}$, whereas the extrinsic contribution will
contain $\delta f$ and hence depend on the scattering processes (manifested,
e.g., by the carrier relaxation time). Clearly, $\Delta \bm M^{\text{int}}$
is of more interest, so below we will focus on the intrinsic contribution.
The extrinsic contribution is analyzed in Refs. \cite{Murakami2015,Zhong2016},
and its typical behavior will be commented later.

Second, to account for $\Delta \bm M$ in the linear order of the driving forces, it is sufficient to take $\mathcal{D}=1$ in
the second term on the right hand side of (\ref{j-loc}), so that the contribution of this term to $\Delta \bm M^{\text{int}}$ is ($\partial
_{j}\equiv \partial /\partial r_{j}$ henceforth) $\boldsymbol{\nabla }\times \int \left[
d\boldsymbol{k}\right] f_{0}[\boldsymbol{\tilde{m}}+(eE_{j}-\partial
_{j}\varepsilon )\partial _{\bm B}a_{j}^{B}]$.

Third, comparing the form of Eq.~(\ref{j-loc}) to Eq.~(\ref{current}), one
might be tempted to identify the second term on the right hand side of (\ref%
{j-loc}) as $\bm j^{\text{mag}}$. However, this is incorrect for $\Delta \bm %
M^{\text{int}}$. Similar to Refs. \cite{Xiao2020EM,Dong2020}, by tracing the
$\bm w$ field, one finds that the first term in (\ref{j-loc}) actually also
contains a part that belongs to $\bm j^{\text{mag}}$. Detailed calculations
\cite{supp} show that {the sum of this part and the $\boldsymbol{\tilde{m}}$
term in Eq.~(\ref{j-loc}) gives the equilibrium OM.}

Finally, according to the above observations, the quantity $(eE_{j}-\partial
_{j}\varepsilon )f_{0}\partial _{\bm B}a_{j}^{B}$ plays the decisive role in
$\Delta \bm M^{\text{int}}$. The statistical forces naturally enter into the
picture through the factor $f_{0}\partial _{j}\varepsilon
=\partial _{j}g_{0}-\partial _{T}g_{0}\partial _{j}T-\partial _{\mu
}g_{0}\partial _{j}\mu $, where $g_{0}=-\int_{\varepsilon }^{\infty }d\eta
f_{0}(\eta )$.

{\color{blue}\textit{Intrinsic orbital magnetoelectric and magnetothermal
effects}.} Combining these considerations and collecting terms that are
linear in the driving forces of $\bm E$, $\nabla T$, and $\nabla \mu $, we
arrive at the following result for the intrinsic field-generated OM:
\begin{equation}
\Delta M^{\text{int}}=\boldsymbol{\chi }^{\text{int}}\cdot (\boldsymbol{E}-{%
\nabla }\mu /e)-\boldsymbol{\alpha }^{\text{int}}\cdot {\nabla }T,
\label{OM}
\end{equation}%
where we have used the fact that for 2D, OM only has the $z$-component (with
$\Delta \bm M^{\text{int}}=\Delta M^{\text{int}}\hat{z}$), and hence it is
convenient to also express the susceptibility tensors in vector forms, given
by
\begin{equation}
\boldsymbol{\chi }^{\text{int}}=e\int \left[ d\boldsymbol{k}\right] f_{0}%
\bm{\Lambda}^{n},  \label{OME}
\end{equation}%
\begin{equation}
\boldsymbol{\alpha }^{\text{int}}=\int \left[ d\boldsymbol{k}\right] s_{0}%
\bm{\Lambda}^{n},  \label{intrinsic OM}
\end{equation}%
with
\begin{equation}
s_{0}=\left( \varepsilon -\mu \right) f_{0}/T+k_{B}\ln [1+e^{-(\varepsilon
-\mu )/k_{B}T}]
\end{equation}%
being the entropy density contributed by a particular state. The common
factor $\bm\Lambda ^{n}$ in Eqs.~(\ref{OME}) and (\ref{intrinsic OM}) is
given by (we restore $\hbar $ and band indices $n$, $m$ here)
\begin{equation}
\Lambda _{i}^{n}=2\operatorname{Re}\sum_{m\neq n}\frac{\mathcal{A}_{i}^{nm}\mathcal{M%
}^{mn}}{\varepsilon _{n}-\varepsilon _{m}}+\frac{e}{2\hbar }\epsilon _{j\ell
}\partial _{k_{j}}\mathcal{G}_{i\ell }^{n},  \label{a-k}
\end{equation}%
where $\bm{\mathcal{A}}^{nm}=\langle u_{n}|i\partial _{\bm k}|u_{m}\rangle $
is the interband Berry connection, $\mathcal{M}^{mn}=(e/2)\sum_{m^{\prime
}\neq n}[(\boldsymbol{v}^{mm^{\prime }}+\boldsymbol{v}^{n}\delta
_{mm^{\prime }})\times \bm{\mathcal{A}}^{m^{\prime }n}]_{z}$ is the
interband orbital magnetic moment with $\boldsymbol{v}^{mm^{\prime }}$ being
the velocity matrix element, $\mathcal{G}_{i\ell }^{n}=\operatorname{Re}\langle
\partial _{k_{i}}u_{n}|\partial _{k_{\ell }}u_{n}\rangle -\bm{\mathcal{A}}%
_{i}^{n}\bm{\mathcal{A}}_{\ell }^{n}$ is the Fubini-Study quantum metric for
band $n$~\cite{QM2020,QM1980}, and $\epsilon _{ij}$ is the 2D Levi-Civita
symbol. Equations (\ref{OM})-(\ref{a-k}) are the key results of this work.
They present for the first time a unified theory for the electrically and
thermally generated OM, applicable for both metals and insulators.

The result exhibits several nice features. First, the first term on the
right hand side of Eq.~(\ref{OM}) explicitly shows that the Einstein
relation, i.e., the equivalence between $\bm E$ field and chemical potential
gradient, is fulfilled in our theory. It has to be stressed that this
fulfillment is quite nontrivial: the $\bm E$ field enters through the
equations of motion (\ref{EOM}) as well as the field corrections in the
quantities in Eq.~(\ref{j-loc}), whereas $\nabla \mu $ enters only through
the occupation function. Therefore, the consistency with the Einstein
relation serves as a nice check for the validity of our theory.

Second, the two susceptibilities $\bm\chi ^{\text{int}}$ and $\bm\alpha ^{%
\text{int}}$, representing magnetoelectric and magnetothermal responses,
satisfy the generalized Mott relation
\begin{equation}
\boldsymbol{\alpha }^{\text{int}}=\int d\eta \ \partial _{\eta }f_{0}\frac{%
\eta -\mu }{eT}\boldsymbol{\chi }^{\text{int}}\left( \eta \right) ,
\end{equation}%
where $\boldsymbol{\chi }^{\text{int}}\left( \eta \right) $ is the value of (%
\ref{OME}) at zero temperature, taken to be a function of the chemical
potential $\eta $. At low temperatures, the above equation reduces to
\begin{equation}
\boldsymbol{\alpha }^{\text{int}}=\frac{\pi ^{2}k_{B}^{2}T}{3e}\partial
_{\eta }\boldsymbol{\chi }^{\text{int}}\left( \eta \right) |_{\eta =\mu }.
\label{Mott}
\end{equation}%
Traditionally, the Mott relation is between the electric and thermal
conductivities. Recent works have extended its applicability to various
Berry curvature related linear responses, such as spin polarization and spin
torques \cite{Freimuth2014,Shitade2019,Dong2020}. Here, we have shown that
it also establishes a connection between orbital magnetoelectric and
magnetothermal effects.

Third, as promised, the intrinsic susceptibilities $\bm\chi ^{\text{int}}$
and $\bm\alpha ^{\text{int}}$ comprise only band geometric quantities.
Interestingly, the Berry curvature does not appear in (\ref{a-k}), instead,
there emerges the quantum metric. Geometrically, the quantum metric measures
the distance between neighboring Bloch states. It has attracted great
interest, because of its appearance in various nonlinear effects discovered
in recent works \cite{QM2020,Lapa2019,Gao2015}. Note that all quantities in (%
\ref{a-k}) are gauge invariant, which makes it convenient to be combined
with first-principles calculations for real materials.

{\color{blue}\textit{Application to bilayer CrI$_{3}$}.} {We demonstrate the
application of our theory in studying a real material. It is important to
note that crystalline symmetries also impose constraints on these effects.
From Eq.~(\ref{OM}), we see that the susceptibilities behave as in-plane
pseudo-vectors (the same symmetry as the Berry curvature dipole~\cite{Fu2015}%
). Thus, the largest spatial symmetry allowed in 2D is a single mirror line.
Guided by this constraint, we consider the example of bilayer CrI$_{3}$.

\begin{figure}[ptb]
\centering
\includegraphics[width=1\columnwidth]{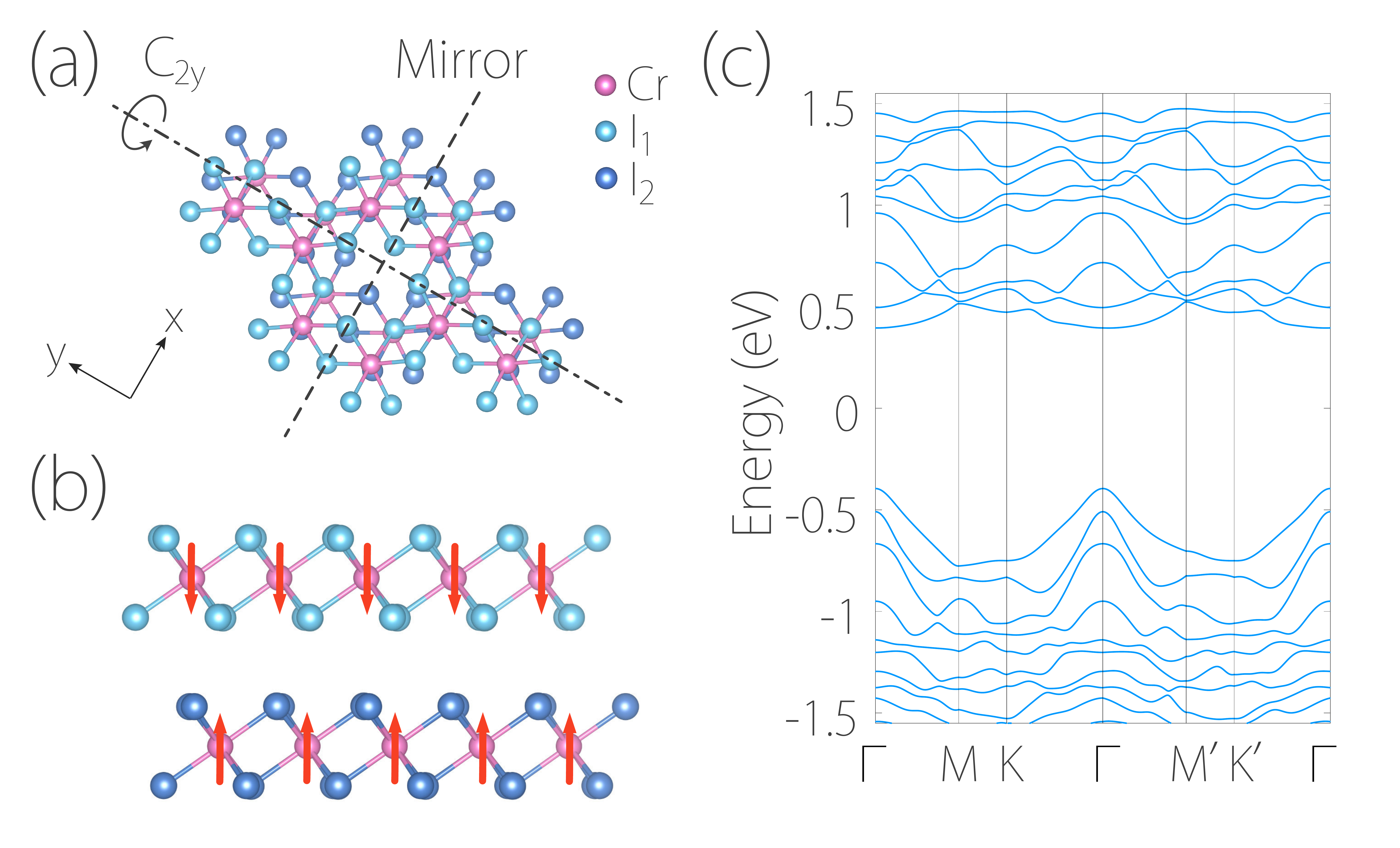}
\caption{(a) Top and (b) side view of the structure of CrI$_{3}$ bilayer. Green (blue) balls represent I atoms in the top (bottom)
layer. The arrows depict the local spin orientation.
(c) Calculated band structure of CrI$_{3}$ bilayer.}
\label{fig:biCrI3}
\end{figure}

\begin{figure}[ptb]
\includegraphics[width=1\columnwidth]{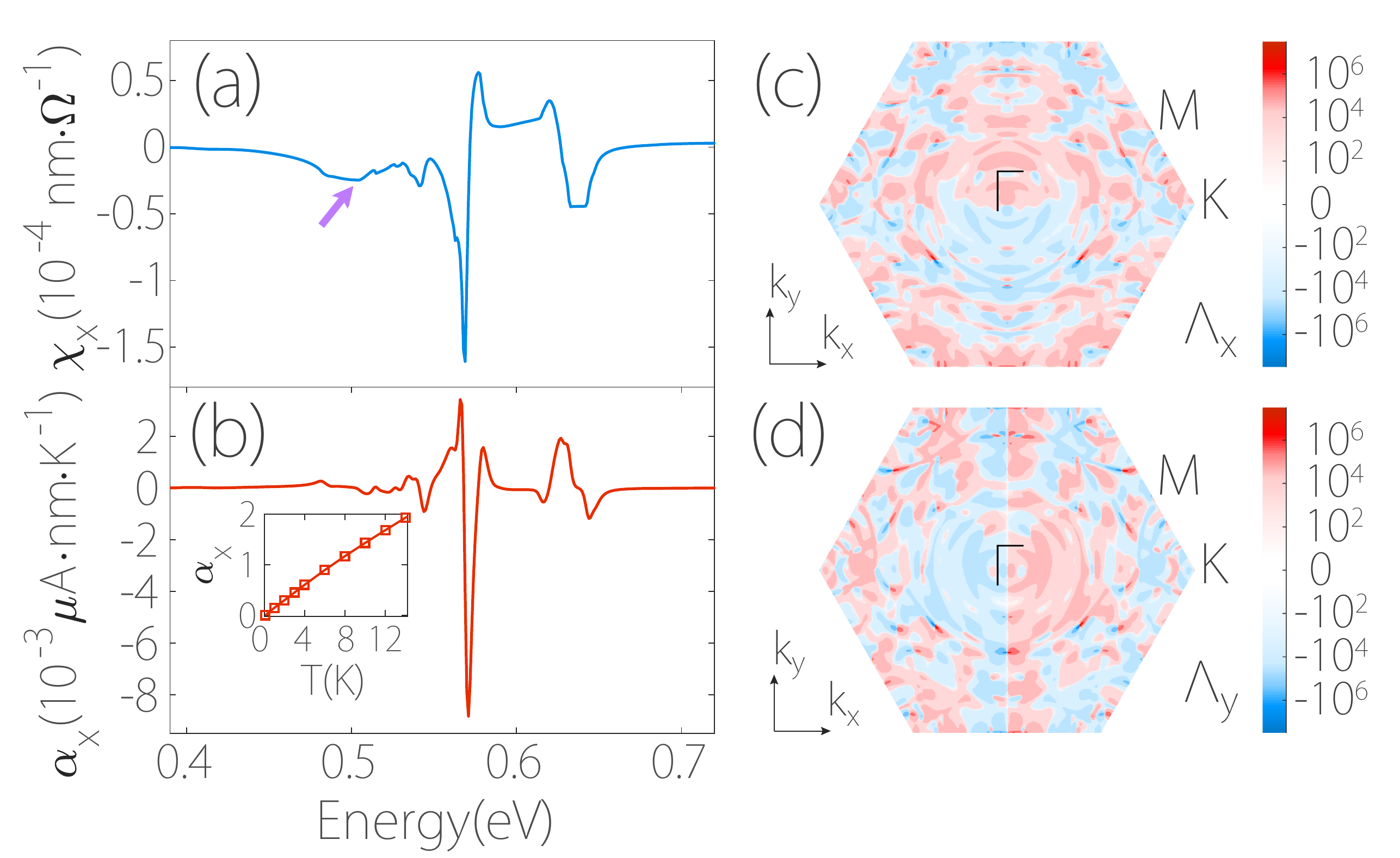} \centering
\caption{Calculated susceptibilities (a) $\chi_x$ and (b) $\alpha_x$ versus the chemical potential for CrI$_3$ bilayer.
The inset in (b) shows the temperature dependence of $\alpha_x$ at 0.5 eV. (c) and (d) show the momentum space distribution of  $\Lambda_{x}(\boldsymbol{k})$ and $\Lambda_{y}(%
\boldsymbol{k})$ at 0.5 eV (marked by the arrow in (a)), in the unit of $%
e\hbar^{-1}\cdot\text{\AA }^{-3}$. In the calculation, we take $T=10$ K.
}
\label{fig:magthem}
\end{figure}

CrI$_{3}$ is a van der Waals layered magnetic material. Monolayer and
bilayer CrI$_{3}$ have been successfully fabricated in experiment. The
bilayer CrI$_{3}$ has a monoclinic ($C2/m$) crystal structure~\cite{Sun2019}%
, and it was demonstrated to be a 2D antiferromagnet at ground state~\cite%
{Huang2017,Seyler2018}. As shown in Figs. \ref{fig:biCrI3}(a) and (b), the
magnetic moments are mainly on the Cr sites. The coupling within each layer
is ferromagnetic, whereas the interlayer coupling is of antiferromagnetic
type. Clearly, both $\mathcal{P}$ and $\mathcal{T}$ are broken in this
system, therefore the intrinsic orbital magnetoelectric and magnetothermal
effects are allowed. Furthermore, the configuration preserves the $\mathcal{%
PT}$ symmetry, under which the intrinsic contribution will be the dominant
OM response (the equilibrium OM also vanishes, see Table~\ref{tab:Basic}).

The susceptibilities $\bm\chi ^{\text{int}}$ and $\bm\alpha ^{\text{int}}$
have been evaluated by our theory combined with DFT calculations (see
Supplemental Material \cite{supp} for details). Figure \ref{fig:biCrI3}(c)
shows the band structure of bilayer CrI$_{3}$ obtained from DFT
calculations. Note that the system has a twofold rotational axis $c_{2y}$
(see Fig. \ref{fig:biCrI3}(a)), which requires $\bm\chi ^{\text{int}}$ and $%
\bm\alpha ^{\text{int}}$ to be along the $x$ direction. This feature is
confirmed by our DFT calculation. In Fig. \ref{fig:magthem}, we plot the
values of the two susceptibilities as functions of the chemical potential,
which can physically be tuned by gating (for simplicity, we fix the magnetic
configuration in the calculation, which in reality may be achieved by
pinning with neighboring magnetic layers). One observes that both $\bm\chi ^{%
\text{int}}$ and $\bm\alpha ^{\text{int}}$ are very small inside the band
gap. However, they increase rapidly under doping. $\chi ^{\text{int}}$ can
reach a typical magnitude of $10^{-4}$ nm/$\Omega $. Assuming an applied $E$
field of $10^{4}$ V/m (along $x$), the induced magnetization, which is
out-of-plane, can reach $\sim 10^{-4}\mu _{B}/\text{nm}^{2}$, which is two
orders larger than that observed in doped monolayer MoS$_{2}$~\cite{Mak2017}%
, and is of the same order as the Edelstein effect in strongly Rashba
spin-orbit-coupled materials such as the Au(111) surface~\cite{Johansson2016}.

The Mott relation in (\ref{Mott}) indicates a linear temperature scaling for
$\alpha ^{\text{int}}$ at low $T$ regime. This behavior is also explicitly
demonstrated by our calculation, as shown in the inset of Fig. \ref%
{fig:magthem}(b). In addition, in Figs. \ref{fig:magthem}(c) and (d), we
plot the $\bm k$-resolved $x$ and $y$ components of $\bm\Lambda (\bm %
k)=\sum_{n}f(\varepsilon _{n})\bm\Lambda ^{n}(\bm k)$, defined for all
occupied states. One observes that governed by the $c_{2y}$ symmetry, $%
\Lambda _{x}$ ($\Lambda _{y}$) is an even (odd) function with respect to the
$y$ axis. It follows that only the $x$ component survives after the
integration over the Brillouin zone.}

{\color{blue}\textit{Discussion}.} In this work, we have predicted a new
effect---the orbital magnetothermal effect, and we have developed a unified
theory for both orbital magnetoelectric and magnetothermal effects. In the
main text, we have focused on the intrinsic contribution, which is solely
determined by the band structure properties, regardless of the carrier
scattering. In general, extrinsic contribution would also exist. For
instance, when $\mathcal{P}$, $\mathcal{T}$ and $\mathcal{PT}$ are all
broken, the extrinsic and intrinsic OM responses may compete with each
other. Nevertheless, we can distinguish them by their different scaling
behavior. We find that when elastic or quasi-elastic scattering dominates,
the extrinsic contributions also comply with the Mott relation $\boldsymbol{\alpha }^{%
\text{ext}}=\frac{\pi ^{2}k_{B}^{2}T}{3e}\partial _{\eta }\boldsymbol{\chi }%
^{\text{ext}}\left( \eta \right) |_{\eta =\mu }$.  In the
high-$T$ regime, the electron-phonon scattering dominates, we have $%
\boldsymbol{\alpha }^{\text{ext}}\sim T^{0}$ due to the $T^{-1}$ scaling of
the electron-phonon relaxation time $\tau _{ep}$ \cite{Ziman1972}. At low
temperatures, the electron-impurity scattering dominates, then $\boldsymbol{%
\alpha }^{\text{ext}}\sim T$, due to the $T$ independence of the
electron-impurity relaxation time $\tau _{i}$.

A temperature gradient can induce an off-equilibrium phonon distribution,
which in turn drives electrons out of equilibrium through the
electron-phonon coupling and hence may generate an additional extrinsic
contribution, the phonon-drag OM. The pertinent correction to the electronic
occupation has the form of $\delta f^{\text{g}}\propto (\tau _{i}\tau
_{p}/\tau _{ep})(\nabla T/T)$ in the relaxation time approximation, where $\tau _{p}$ is the phonon relaxation time. At low
temperatures, $\tau _{p}$ is a constant, limited by the boundary scattering~%
\cite{Lyo1988}, and $\tau _{ep}^{-1}$ usually takes the form of a power-law,
e.g., $T^{4}$ in a MoS$_{2}$ monolayer~\cite{Jauho2013}. The phonon-drag OM
thus increases with $T$ until the phonon-phonon scattering degrades $\tau
_{p}$ significantly. Therefore, a peak in its $T$ dependence is anticipated,
similar to the well-confirmed peak in the phonon-drag thermopower \cite%
{Lei2010}.

Finally, we mention that in doped ionic materials with reduced symmetry,
under a temperature gradient, the off-equilibrium phonons carrying angular
momentum~\cite{Zhang2014,Spaldin2019} may also produce an induced OM~\cite%
{Murakami2018}. Being proportional to $\tau _{p}$, this contribution is
expected to be strongly suppressed at high temperatures, where the electron
contribution prevails. Besides, it scales as $T^{2}$ at low temperatures in
2D \cite{Murakami2020}, which is also sub-dominant compared to the electron
contribution $\sim T$. This qualitative analysis suggests that the electron
OM is likely to be dominating in a wide temperature regime.

\begin{acknowledgments}
We thank B. Xiong, L. Dong, Y. Gao and T. Chai for useful discussions.
C.X. and Q.N. are supported by NSF (EFMA-1641101) and Welch Foundation (F-1255). H.L., J.Z. and S.A.Y. are supported by
the Singapore MOE AcRF Tier 2 (MOE2017-T2-2-108).
\end{acknowledgments}


\begin{thebibliography}{99}
\bibitem{note-uncertainty} From Eq.~(\ref{current}), OM can only be
determined up to a gradient in 3D (there is not such an uncertainty in 2D).
In equilibrium, OM could be a well behaved bulk property. However, under $%
\bm E$ field, the definition must involve the electric polarization, which is
ill defined in metals. For 3D metals, it is currently unknown
whether the field-induced OM can be well defined.
\end{thebibliography}
\end{document}